\documentstyle{article}

\def\6{\langle}
\def\9{\rangle}
\def\half{\mbox{$1\over2$}}

\begin{document}

\title{What is actually teleported?}

\author{Asher Peres\thanks{Department of Physics, Technion---Israel
Institute of Technology, 32000 Haifa, Israel}}

\maketitle
\begin{abstract}
There are no ``unknown quantum states.'' It's a contradiction in terms.
Moreover, Alice and Bob are only inanimate objects. They know nothing.
What is teleported instantaneously from one system (Alice) to another
system (Bob) is the applicability of the preparer's knowledge to
the state of a particular qubit in these systems. The operation
necessitates dual classical and quantum channels.  Other examples of
dual transmission, including ``unspeakable information,'' will be
presented and discussed. This article also includes a narrative of
how I remember that quantum teleportation was conceived.

\end{abstract}

\bigskip

\section{Birthdays}
It is a great pleasure to participate in the IBM Symposium honoring
the 60th birthday of Charles Bennett. I knew some of Bennett's works as
soon as I became interested in quantum information when I visited John
Wheeler in Austin in 1979. However I actually met Charlie only in the
summer of 1986, when I spent two months at MIT. We both lived in the
house of Tom Toffoli, who also was our host at MIT. Tom had bought a
dilapidated house in Howard Street and was busy making it livable. His
family had the third floor, I was in the second floor in a tiny
apartment that was perfect for me, and Charlie, Theo, and her children
were in a larger apartment, also in the second floor. The ground floor
had not yet been rebuilt and looked like a construction site.

This time is also the 10th anniversary of quantum teleportation, an
article that I had the honor of co-authoring with Charles Bennett,
Gilles Brassard, Claude Cr\'epeau, Richard Jozsa, and William
Wootters \cite{telep}.
I shall discuss only the title of that
paper, ``Teleporting an Unknown Quantum State via Dual Classical
and Einstein-Podolsky-Rosen Channels'' (there is no time for more
than the title), and I'll relate what I remember of how this work
was conceived. I apologize if my memory failed in some cases, or I
unwittingly distorted the truth.

In October 1992, Bill Wootters (whom I knew from Austin, where he
had been a student), sent me an e-mail saying that he and others
in Montreal had found an interesting problem, and he asked for my
advice. When things became clearer and we thought of writing a paper
with six co-authors, we started arguing on every nuance of the text. All
this had to be done by e-mail, because we were then scattered in five
different places in four countries and eight time zones. Some of us
worked while others were sleeping. Charlie quipped ``the Sun never
sets on our collaboration'' and thereby started an argument who was
the king who had said that. First we thought of Charles Quint, but
after some research work it turned out that it had been Phillip~II.

There were some memorable moments while the text was finalized. One
Friday afternoon, Claude sent me an e-mail from Paris: what happens if
Alice's particle, whose state has to be teleported, is itself entangled
with another one, far away? Will Bob's particle become entangled with
this other particle, without having ever interacted with it? I was
puzzled, but it was time to start the traditional Shabbat diner
with my family. As we were eating, I suddenly jumped from my seat, ran
to the computer and wrote to Claude ``mais oui.'' He had invented
entanglement swapping! 

Charlie did most of the editing. When everything looked fine, I sent
him an e-mail with subject: {\it imprimatur\/} (the seal of approval of the
Great Inquisitor). Charlie submitted the paper to PRL, and wrote to us
{\it alea jacta est\/}, as if we had crossed the Rubicon. Contrary to
expectations, our opus was not rejected by the referees. Later we learnt
that one of them was David Mermin who gave a very strong recommendation 
that it had to be published. It's only more recently that David 
deconstructed teleportation, and also dense coding \cite{mermin}.

Not only the contents of the teleportation paper are interesting,
but also what is {\it not\/} in it. There are no acknowledgments for
support by NSF, NASA, DARPA, NRL, and other philanthropic agencies. We
never submitted a research proposal, that would have been rejected
anyway. There was no time for that.\footnote{My last research
proposal, about 25 years ago, was rejected by BSF as being a ``high
risk project.'' I asked ``Which risk? I am risking to waste my time,
what are you risking?'' The BSF representative explained that this
research might not have the expected results. They wanted to be sure
that I'll write a report with all the answers to the questions I had
raised.} Now, let's start and analyze the title of the paper.

\section{Teleporting}

I don't watch TV and I was suspicious of the term teleportation.
In my dictionary \cite{webster},
I found ``theoretical
transportation of matter through space by converting it into energy and
then reconverting it at the terminal point.'' I protested that this was
not at all what we had in mind, but Charlie reassured me, saying that
we shall cite Penrose's book. I threatened that if we cite Penrose,
I won't be a co-author. A few days later, Charlie wrote to me that he
wanted to use weak measurements and cite Aharonov and Vaidman. This
time, I didn't fall in the trap.

We had other semantic problems. I proposed to write that the quantum
state was disembodied and reincarnated. This was found unacceptable.
Later, when a newsman asked me whether it was possible to teleport not
only the body but also the soul, I answered ``only the soul.'' Even
that is a gross oversimplification.

\section{Unknown quantum state}

The notion ``quantum state'' encapsulates what is {\it known\/}
of the preparation of a system \cite{whatis}.  An unknown quantum
state is a contradiction in terms, an oxymoron, just as a ``research
proposal.'' Enrico Fermi said that when there is no surprise, it's
not research.

Anyway, Alice and Bob are not real people. They are inanimate objects.
I have seen an optical bench with a label {\sc alice} near a piece of
hardware, and {\sc bob} near another one. The hardware knows nothing.
What is teleported instantaneously from one system (Alice) to another
system (Bob) is the applicability  of the preparer's knowledge to
the state of a particular qubit in these systems \cite{opinion}. The
preparer whose knowlegde is teleported is a real person with a PhD
in physics. His name is Chris.

The next item in the title are the dual classical and EPR channels.
The text we submitted said ``EPR'' and the APS editorial office
automatically expanded this acronym into ``electron paramagnetic
resonance.'' Somebody caught the error and restored the dignity of
Einstein, Podolsky and Rosen.\footnote{This does not always
happen. See Phys.\ Rev.\ A {\bf64}, 042310 (2001), line 12 of text.}
Dual classical and quantum channels still are an open problem, and 
I'll keep them for the end.

\section{Quantum archaelogy}

``The discovery of quantum teleportation grew out of an attempt
to identify what other resource, besides actually being in the same
place, would enable Alice and Bob to make an optimal measurement of the
Peres-Wootters states.'' \cite{mor}
In 1980, during my second visit to John
Wheeler at Austin, I shared an office with Bill Wootters who had just
submitted his Ph.D. thesis ``The acquisition of information from quantum
measurements.'' In that thesis, there were two observers, the ancestors
of Alice and Bob, who used polarized photons to communicate quantum
information. Soon after that, I read a fascinating article ``Unforgeable
Subway Tokens'' \cite{tokens}
that I had found during a bibliographic search of Charlie's
works, because I was interested in the thermodynamics of information.

In 1989, the Santa Fe Institute organized a workshop on complexity,
entropy, and the physics of information. Bill was there on sabbatical
leave, and I stayed an extra couple of weeks in order to work with him.
We discussed the following problem: given two quantum systems in the
same state, can we acquire more information by a joint measurement on
both, than by separate measurements on each one, assisted by classical
communication (the acronym LOCC didn't exist yet). My intuition was that
a joint measurement would in some cases be more efficient, and Bill's
intuition was the opposite. I proposed a few simple examples, for which
Bill showed that his opinion was correct. As I had to leave SFI, we
decided to continue by using bitnet.

\begin{figure} \vspace{1in}
\caption{sfi.jpg will be supplied separately}
\end{figure}

\section{BITNET}

Before e-mail, there was BITNET (``because it's time net'') a
service that was provided gratis by IBM for a few years. IBM is not
a philanthropic institution. The strategy was similar to that of drug
pushers who offer free cocaine to children. After the kids are hooked,
they need the drug and pay dearly for it. The difference is that when
IBM discontinued bitnet, it was replaced by another free service,
Internet (I don't know who actually pays for that).

I learnt of bitnet in 1985, when Murray Peshkin at ANL wanted to
communicate with me and asked Harry Lipkin at Weizmann what was
my address. Harry explained to me the theory, and soon after that
Murray sent me a first message: ``Welcome to the brave new world of
bitnet!'' Likewise I taught the magic to Bill and welcomed him in the
brave new world. All this was quite primitive by today's standards, with
a 1200 baud modem. After a few other unsuccessful attempts to prove to
Bill that joint measurements could be more efficient, I proposed trine
states,\footnote{A trine is an astrological configuration where three
planets make angles of $120^\circ$. The word trine was introduced by
Charlie, because I disliked Mercedes and no one would take Mitsubishi.}
with the property that
$$\6\psi_1,\psi_2\9\6\psi_2,\psi_3\9\6\psi_3,\psi_1\9=-\mbox{$1\over8$}.$$
This is the most negative number that can be obtained with any three
states. For example, photons linearly polarized $2\pi/3$ apart, or
spin-\half\ particles polarized $4\pi/3$ apart, form a trine (note
that fermions have to be rotated by $4\pi$ to return to the original
state). This was a lucky guess. It was recently proved \cite{jozsa}
that a trine measurement has the largest entanglement cost of all POVMs.

With a pair of identical trine states, it was impossible to match
the mutual information obtainable from a joint measurement by means
of a small number of LOCC steps, and Bill devised a ``ping-pong''
method with a sequence of POVMs, converging to some optimum. These
were long and difficult calculations. Bill used a MacIntosh with {\sc
pascal}. I had an IBM PS/2 with {\sc fortran}. When our results agreed,
we were pretty sure that there was no numerical error. The optimal
mutual information that we could obtain in this way was less than that
of a joint measurement. On 15 February 1990, we submitted our paper
\cite{pw91} and naturally ran into trouble with the referees. The
typical reaction was: it may be correct, but why is this interesting?
As I tried to explain the paper to one of my colleagues at Technion,
he quipped with a grimace ``it's only engineering.'' Our paper was
thus rejected by PRL and I had to convince a reluctant Bill to appeal
to the Editorial Board. Our appeal was adjudicated by Tony Leggett
and our opus finally appeared on 4 March 1991.

\section{Meeting all the teleporters}

In October 1992, there was in Dallas a meeting on physics and
computation. I introduced Bill Wootters to Charlie, and told him of our
work. Charlie already knew it. He pulled a copy from his briefcase,
and told us that he was showing it to everybody. Later he introduced
me to Gilles Brassard and we immediately were friends, as we could
speak French.  I also met for the first time Richard Jozsa, who was
at that time in Montreal, and Claude Cr\'epeau who was then based
in Paris. Gilles invited Bill to give a seminar at Universit\'e de
Montr\'eal. Everybody but me was there. After the seminar, there was
a discussion in Gilles's office and the question was raised what other
resource would enable Alice and Bob, far away from each other, to make
an optimal measurement of the trine states. After everyone returned
home, Bill sent me a bitnet. I already told the rest of the story.

\begin{figure} \vspace{1in}
\caption{The first Torino workshop on quantum information. From left to
right, first row: Massimo Palma, Claude Cr\'epeau, Gilles Brassard,
Charles Bennett, Bruno Huttner, Umesh Vazirani, David Deutsch; second
row: Mai-Mai Lam, Artur Ekert, Andr\'e Berthiaume, Wojtek Zurek, Asher
Peres, Neil Gershenfeld, Bill Wootters, Mario Rasetti, Roger Penrose;
third row: Ben Schumacher, Carl Caves, Juan-Pablo Paz, G\"unter Mahler,
Andy Albrecht, Richard Jozsa, Norman Margolus, Giuseppe Castagnoli.}
\end{figure}

In June 1993, there was the first Torino workshop on quantum
information. Today, there are hundreds of participants in quantum
information conferences, but at that time we were only a small number
of addicts (Fig.~2). The two gentlemen with bizarre dresses are
the most important people: one of them collects money for ISI (Institute
for Information Interchange, in Torino) and the other one spends that
money and organizes meetings. Everyone in that picture is still active
in the field, except Mai-Mai Lam who chose a different career, a real
loss for the quantum information community.

\section{Group picture}

The weather in Villa Gualino was wonderful. Claude lent his camera to
Andr\'e Berthiaume who took a group picture of the six teleporters. When
Claude returned to Paris, he arranged to have an article on quantum
teleportation appear in the popular scientific magazine {\it Science et
Vie\/} \cite{scvie}. The following month, I was a few days in Tournai
and bought the journal in a newstand, but not before I checked that my
picture was indeed in it. The newstand owner was flabbergasted.

\begin{figure} \vspace{1in}
\caption{sc-vie.jpg will be supplied separately}
\end{figure}

Our next group picture, with exactly the same configuration, was taken
twice in Cambridge (UK) in July 1999. Not far from us, there was a
big cat, and Charlie later manipulated the photos so that the cat
(whom he called {\it teleportus\/}) appeared distorted in the first
picture, but properly Pauli rotated in the second one. The true name
of the cat was Sam \cite{lei}.

\begin{figure} \vspace{1in}
\caption{Cambridge double picture to be supplied by Charlie}
\end{figure}

\section{Dual classical and quantum channels}

Dual classical and quantum channels have a long history in quantum
information theory. In the classic BB84 protocol \cite{bb84}, each
successful attempt of Alice and Bob to produce a random {\it secret\/} 
bit shared by both of them costs one qubit and two public bits of 
classical information. In the teleportation protocol \cite{telep}, the
remote preparation of one qubit requires one EPR pair, one local qubit,
and two bits of public information.

Dual channels are also needed for ``unspeakable'' quantum information,
namely information that cannot be represented by a sequence of discrete
symbols. For example, Alice wants to indicate to Bob a direction
in space.  If they have a common coordinate system to which they can
refer, or if they can create one by observing distant fixed stars,
Alice simply communicates to Bob the components of a unit vector
{\bf n} along that direction, or its spherical coordinates $\theta$
and $\phi$. But if no common coordinate system has been established,
all she can do is to send a real physical object, such as a gyroscope,
whose orientation is deemed stable.

In the quantum world, the role of the gyroscope is played by a system
with large angular momentum.  The fidelity of the transmission
is usually defined as $$ F=\6\cos^2(\chi/2)\9=(1+\6\cos\chi\9)/2, $$
where $\chi$ is the angle between the true {\bf n} and the direction
indicated by Bob's measurement. The physical meaning of $F$ is that the
{\it infidelity\/} $1-F=\6\sin^2(\chi/2)\9$ is the mean square error
of the measurement \cite{petra}. The experimenter's aim, minimizing
the mean square error, is the same as maximizing fidelity.

Massar and Popescu \cite{mp} took $N$ parallel spins, polarized
along~{\bf n}, and showed that $1-F=1/(N+2)$. It then came as
a surprise that for $N=2$, parallel spins were not the optimal
signal, and a slightly higher fidelity resulted from the use of
opposite spins~\cite{gp}. This better result also required, of
course, the transmission of a classical bit, to tell which spin
was parallel and which one opposite to {\bf n}. This raised the
question what was the most efficient signal state for $N$ spins.
How quickly will $F$ tend to~1? Peres and Scudo \cite{petra} and
a Barcelona group \cite{catalans} showed that the optimal result
was a quadratic approach, as illustrated in Fig.~5. This, however
necessitates that the $N$ spins be distinguishable (for example, a
proton, an electron, and so on). Then there are $N!$ possible ways
of labelling these $N$ spins, requiring the transmission of about
$N\log_2N$ classical bits.

\begin{figure} \vspace{1in}
\caption{Transmission of a direction by $N$ spins. Open circles are
the results of Massar and Popescu \cite{mp}, closed circles are those
of Peres and Scudo \cite{petra}.}
\end{figure}

\section{EPR vs quantum teleportation}

In the EPR-Bohm scenario \cite{bohm}, Alice and Bob share a pair of
spin-\half particles in a singlet state. Alice measures a component of
her spin, and then she knows {\it instantaneously\/} the corresponding
component of Bob's spin \cite{interv2}, namely Bob's result if he
measures (has measured, will measure) the same component of his spin.
However, Alice cannot choose the result she obtains.

In the teleportation scenario, Chris chooses the state of the qubit he
prepares near the apparatus called {\sc alice}. He also prepares an
EPR pair and places the two entangled spins with {\sc alice} and {\sc
bob}, far away from each other. Then {\sc alice} is used to measure
the two spins in her location so that Chris knows which one of the
four possible results was obtained; and then Chris {\it immediately\/}
knows the state of the spin located at {\sc bob}.

The scenario could stop here. There is no compelling reason to transmit
the two classical bits to {\sc bob}, if we are satisfied with a rotated
{\it teleportus\/} as in Fig.~4a. But then the process would not have
been called teleportation and attracted so much attention \ldots


\begin{thebibliography}{99}

\bibitem{telep}C. H. Bennett {\it et al.\/}, Phys. Rev. Letters,
{\bf70}, 1895 (1993).

\bibitem{mermin}N. D. Mermin, Phys. Rev. A {\bf65}, 012320 (2002);
{\it ibid.\/} {\bf66}, 032308 (2002).

\bibitem{webster}{\it Webster's New World Dictionary of the
American Language\/} (Collins, New York, 1974).

\bibitem{whatis}A. Peres, Am. J. Phys. {\bf52}, 644 (1984).

\bibitem{opinion}C. A. Fuchs and A. Peres, Physics Today
{\bf53} (3) 70 (2000).

\bibitem{mor}C. H. Bennett {\it et al.\/}, Phys. Rev. A {\bf59}, 1070
(1999).

\bibitem{tokens}C. H. Bennett {\it et al.\/}, {\it Advances in
Cryptology\/} (Proceedings of Crypto-82, Plenum, New York, 1983) p.~267.

\bibitem{jozsa}R. Jozsa {\it et al.\/}, quant-ph/0303167. 

\bibitem{pw91}A. Peres and W. K. Wootters, Phys. Rev. Lett. {\bf66},
1119 (1991).

\bibitem{scvie}H. Guillemot, Science et Vie, no. {\bf910}, 40 (1993).

\bibitem{lei}Lei Poo, private communication (1999).

\bibitem{bb84}C.H. Bennett and G. Brassard, in {\it Proceedings of IEEE
International Conference on Computers, Systems and Signal Processing,
Bangalore, India\/} (IEEE, New York, 1984) p.~175.

\bibitem{petra}A. Peres and P. F. Scudo, Phys. Rev. Lett. {\bf 86}, 4106
(2001).

\bibitem{mp}S. Massar and S. Popescu, Phys. Rev. Lett. {\bf74}, 1259
(1995).

\bibitem{gp}N. Gisin and S. Popescu, Phys. Rev. Lett. {\bf83}, 432
(1999).

\bibitem{catalans} E. Bagan, M. Baig, A. Brey, R.  Mu\~noz-Tapia,
and R. Tarrach, Phys. Rev. A {\bf 63}, 052309 (2001).

\bibitem{bohm}D. Bohm, {\it Quantum Theory\/} (Prentice-Hall, New York,
1951) p.~614.

\bibitem{interv2}A. Peres, Phys. Rev. A {\bf61}, 022117 (2000).

\end{thebibliography}
\end{document}